\documentclass{article}

\usepackage[preprint]{neurips_2026}
\usepackage{amsmath}
\usepackage{adjustbox}
\usepackage{multirow} 
\usepackage{xcolor}

\usepackage[utf8]{inputenc} 
\usepackage[T1]{fontenc}    
\usepackage{hyperref}       
\usepackage{url}            
\usepackage{booktabs}       
\usepackage{amsfonts}       
\usepackage{nicefrac}       
\usepackage{microtype}      
\usepackage{xcolor}         

\newcommand{\tempresult}[1]{\textcolor{black}{#1}}
\newcommand{\tempresultself}[1]{\textcolor{black}{#1}}
\title{Rethinking PPG-based Sleep Staging: \\
Datasets, Metrics, and Benchmarks}

\author{
  Shuntian Zheng$^{1}$ 
  Jiawei Wang$^{2}$ 
  Cong Fu$^{3}$ 
  Huan Yu$^{3, *}$ 
  Chen Chen$^{4, *}$ 
  Yu Guan$^{1, *}$ 
  Sai Gu$^{2}$ \\
  $^1$ Department of Computer Science, The University of Warwick \\
  $^2$ School of Engineering, The University of Warwick \\
  $^3$ Department of Neurology, Huashan Hospital, Fudan University \\
  $^4$ Human Phenome Institute, Fudan University \\
    \texttt{shuntian.zheng@warwick.ac.uk}~
    \texttt{Davy.Wang@warwick.ac.uk}~~~ \\
    \texttt{fucong@fudan.edu.cn}~~~
    \texttt{dr.yuhuan@163.com}~~~
    \texttt{chenchen\_fd@fudan.edu.cn}~~~\\
    \texttt{yu.guan@warwick.ac.uk}~~~
    \texttt{Sai.Gu@warwick.ac.uk} \\
    $^*$ Co-corresponding authors
}

\begin{document}

\maketitle

\begin{abstract}
Automated sleep staging assigns discrete stage labels to successive time epochs throughout an overnight recording; conventionally each window spans at least 30 seconds, reflecting the minimum temporal resolution of the clinical scoring standard.
Wearable photoplethysmography (PPG) has attracted sustained interest as an ambulatory alternative to laboratory-based polysomnography, which relies on electroencephalography (EEG) and other recording modalities that are impractical outside clinical environments.
Yet PPG-based staging trails EEG-based methods by a substantial margin, and we argue this gap largely reflects a mismatch between signal and task.
Within a stable stage, PPG's inter-stage feature differences are more subtle than those in EEG; yet at stage boundaries, PPG's principal cardiovascular features, heart rate variability and pulse morphology, shift sharply within seconds.
The conventional practice of assigning one label to each 30-second epoch therefore suppresses feature that is concentrated near boundaries.
We address this gap in two steps.
First, we develop a label expansion pipeline based on Hidden Semi-Markov Models that converts coarse epoch labels into sec-level annotations.
To assess whether these expanded labels are reliable enough for downstream supervision, we validate them on a separate expert-reviewed dataset and through an auxiliary sleep-wake task whose labels are independent of the expansion pipeline.
Second, we use the resulting sec-level supervision on MESA to improve conventional four-class epoch-level staging across four architecturally diverse baselines by 3.7--5.7\,pp in accuracy against the original epoch labels, with supplementary zero-shot evaluation on CFS showing that the transfer benefit persists under cohort and annotation-protocol shift.
%
These results indicate that boundary-aware supervision provides a practical route toward more fully exploiting PPG's distinctive information for sleep staging and supports further evaluation in ambulatory monitoring settings where EEG-based assessment is impractical.

\end{abstract}

\section{Introduction}\label{sec:intro}

Automated sleep staging is a cornerstone of sleep medicine, enabling diagnosis of disorders such as sleep apnea, assessment of circadian rhythm disruption, and evaluation of therapeutic interventions~\cite{ting2005disorders, sateia2014international}.
Standard scoring manuals define five discrete stages, Wake, N1, N2, N3, and rapid eye movement (REM), each assigned to successive 30-second epochs as the minimum temporal unit of clinical annotation~\cite{berry2012aasm}.
The clinical gold standard, polysomnography (PSG), supports sleep staging through simultaneous recording of electroencephalography (EEG), electrocardiography (ECG), and electromyography (EMG), but its dependence on laboratory infrastructure and trained technicians makes large-scale or longitudinal deployment impractical~\cite{markov2025interpretable,berry2012aasm,Zhai20,sleepAI_review2020}.
Wearable photoplethysmography (PPG) has attracted substantial research interest as a low-cost alternative: a single optical sensor embedded in a commercial wearable device continuously measures volumetric blood pulsation, requiring no electrodes and enabling ambulatory use over extended periods~\cite{quino2024optimizing, morokuma2023deep}.
In many ambulatory, home-monitoring, and resource-limited settings where EEG electrode attachment is impractical, wearable optical sensing can provide a practical source of overnight cardiovascular dynamics.

%

Despite this practical relevance, reported performance for PPG-based sleep staging still lags behind EEG-based staging, and this gap has persisted across years of architectural innovation~\cite{wang2025improving}. We argue that the ceiling remains largely unreached not because PPG is devoid of useful physiological information, but because the prevailing task formulation is structurally misaligned with the signal's \textbf{distinctive boundary-sensitivity advantage}.

\begin{figure}[t]
  \centering
  \includegraphics[width=\linewidth]{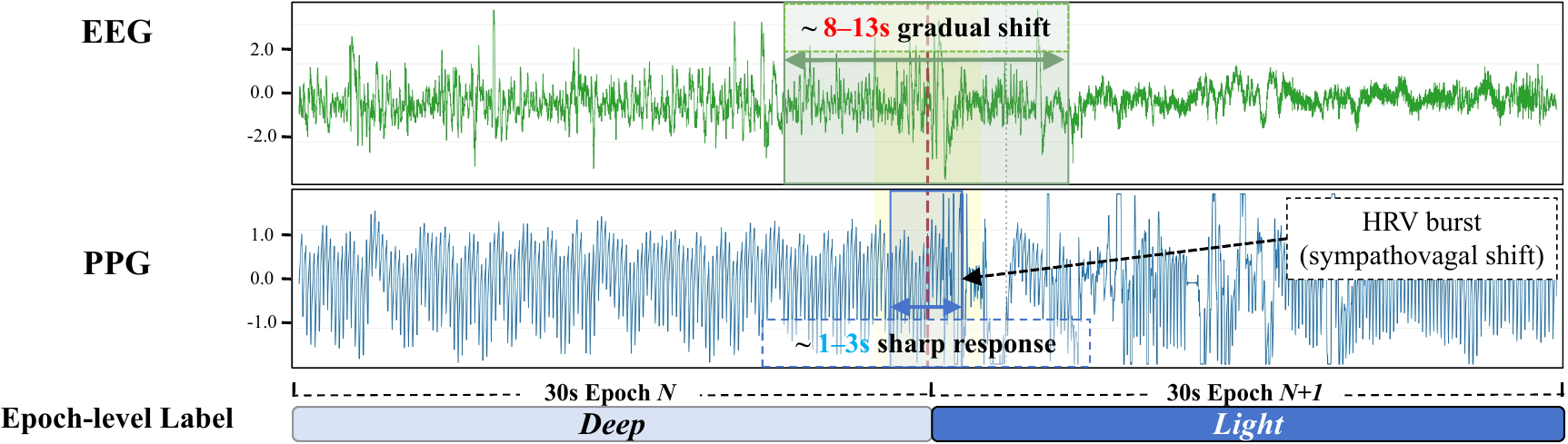}
\caption{Illustrative EEG and PPG signals from continuous sleep segments.
EEG exhibits clearer stage-specific features within stable segments but shifts gradually across boundaries over several seconds, whereas PPG pulses remain more similar across stages yet respond abruptly.
Population-level evidence for the same asymmetry is reported in Section~\ref{subsec:evidence}, Figure~\ref{fig:signal_compare}, and Table~\ref{tab:sensitivity}.
}
  \label{fig:tease} 
\end{figure}

EEG measures cortical synchrony directly and registers sleep stages as spectrally stable patterns that persist across tens of seconds, a property that aligns naturally with the 30-second epoch classification framework used in standard scoring manuals~\cite{berry2012aasm}.
PPG, by contrast, is an indirect cardiovascular signal whose sleep-related content arises from autonomic modulation of heart rate and peripheral vascular tone~\cite{habib2022performance}.
Consistent findings in the physiological literature~\cite{ryals2023photoplethysmography,trinder2001autonomic} are that PPG exhibits pronounced sensitivity to sleep stage transitions: heart rate variability (HRV), the beat-to-beat fluctuation in cardiac inter-beat intervals reflecting sympathovagal balance~\cite{malik1996heart}, and pulse waveform morphology shift sharply at the sec-level timescale around stage boundaries, with autonomic responses emerging within one to two heartbeats and producing a transient perturbation that contrasts sharply against PPG's otherwise stable baseline.
EEG spectral power also shifts at transitions, but does so gradually over several seconds within an already spectrally rich and variable baseline.

Figure~\ref{fig:tease} provides an example of this contrast: the EEG trace changes comparatively smoothly across the true transition onset, while the PPG trace exhibits a narrower boundary-centered perturbation that is not well aligned with the coarser 30-second epoch framing.
Consequently, PPG's boundary response stands out more prominently relative to its own stable-epoch baseline than the corresponding EEG response does, making it more readily detectable under HRV-based and morphological feature representations.
Within stable sleep epochs, however, inter-stage differences in PPG's sustained characteristics are more subtle than the corresponding EEG differences, making the epoch-level static discrimination task intrinsically harder for this modality~\cite{habib2022performance}.
Taken together, these observations reveal a \textbf{characteristic asymmetry}: the physiological advantage of PPG is concentrated at the temporal boundaries between sleep states rather than within the interiors of stable epochs.


%

Despite this characteristic asymmetry, nearly all PPG sleep staging approaches inherit the epoch-level classification paradigm originally developed for EEG and apply it unmodified to PPG~\cite{attia2024sleepppg,kotzen2022sleepppg,fonseca2023computationally, huttunen2021assessment}, leaving the signal's boundary-sensitivity advantage outside the training and evaluation loop. 
Reaching PPG's information-theoretic ceiling for sleep analysis requires a task formulation explicitly aligned with the signal's physiology, yet two structural gaps prevent the community from doing so.
(1) a task-level mismatch: by adopting an epoch classification paradigm originally designed around EEG's within-epoch spectral stability, existing methods implicitly treat PPG as a degraded EEG surrogate.
The transition-sensitive dynamics that PPG encodes most strongly receive no explicit supervision, and evaluation metrics that average over each epoch are blind to boundary localization accuracy.
(2) an infrastructure absence: because no method has targeted transitions, no publicly available dataset provides sec-level sleep stage annotations for PPG, and no established evaluation protocol exists for transition detection.
This absence creates a self-reinforcing cycle in which the characteristic asymmetry remains invisible to both training and evaluation.

This work addresses both traps by proposing Sleep Stage Transition Detection (SSTD) as a complementary task formulation that targets the sec-level onset time and type of each sleep stage transition, the temporal resolution at which PPG's autonomic dynamics are most informative.
To make SSTD tractable, we develop a label expansion framework that converts coarse 30-second epoch labels from existing datasets into sec-level pseudo-labels under a four-class vocabulary of Wake, Light, Deep, and REM, combining Hidden Semi-Markov Models for duration-constrained state inference with local changepoint refinement under physiological transition constraints.
%
We additionally introduce boundary error and transition F1 as evaluation metrics that quantify temporal localization precision and event-level detection completeness with respect to the expanded labels, while separately validating their clinical correspondence through independent physician review.
The separate expert-reviewed dataset is used for this label-fidelity validation; the downstream public-dataset experiments use MESA as the main training and evaluation cohort and CFS as a supplementary zero-shot transfer cohort.



The contributions of this work are as follows.

(1) We identify and empirically verify a characteristic asymmetry in PPG across large-scale PSG-annotated cohorts: under HRV and morphology features, boundary-to-stable contrast ratio substantially exceeds that of EEG, while within-epoch inter-stage separability is markedly lower.
We show that the prevailing epoch-level training paradigm fails to exploit this property, giving rise to \textbf{two interconnected traps} that systematically suppress PPG's potential for sleep staging: a task-level mismatch that treats PPG as a degraded EEG surrogate, and an infrastructure absence that leaves the community without sec-level annotations or task-specific evaluation metrics.

(2) We address both traps by proposing Sleep Stage Transition Detection (SSTD), a task formulation operating at sec-level temporal resolution, together with a physiologically constrained label expansion pipeline for constructing SSTD annotations from existing PSG-annotated cohorts and evaluation metrics for temporal localization.


(3) To validate the pipeline, we collect a separate expert-reviewed PPG dataset with sec-level sleep stage boundaries. Agreement with physician consensus reaches $\kappa = 0.81$ in the adopted four-class setting, supporting the clinical plausibility of the expanded labels. Using these annotations, sec-level supervision improves conventional four-class epoch-level staging on MESA by 3.7--5.7\,pp in accuracy across representative baselines, and the gains persist under zero-shot transfer to CFS, providing a practical basis for improving PPG-based sleep staging in home, longitudinal, and population-scale monitoring settings where EEG-based assessment is impractical or unavailable.


\section{Related Work}\label{sec:related}


\subsection{PPG-Based Sleep Staging}

Prior PPG sleep staging methods span hand-crafted HRV classifiers~\cite{radha2019sleep,fonseca2015sleep}, end-to-end CNN/RNN or spectrogram models~\cite{yildirim2019deep,chambon2018deep,sors2018convolutional,zhao2021multi}, and more recent context-aware or multi-modal architectures~\cite{eldele2021attention,phan2023seqsleepnet,wang2025improving,supratak2017deepsleepnet,constantin2025towards, korkalainen2020deep,coon2025getting}.
Recent work has also explored adjacent directions such as transfer or pre-training, interpretability or uncertainty-aware analysis, multimodal fusion, and wearable-oriented adaptation to improve epoch-level PPG staging~\cite{eldele2021attention,phan2023seqsleepnet,wang2025improving,supratak2017deepsleepnet,constantin2025towards,korkalainen2020deep,coon2025getting,markov2025interpretable}. These lines make clear that the field is actively improving PPG sleep staging along multiple fronts, including representation quality, model inspection, sensor complementarity, and practical deployment analysis.
Our claim is therefore not that prior work ignores PPG-specific modeling, but that these advances are still evaluated predominantly within the inherited 30-second staging setup. Under that formulation, datasets remain epoch-labeled, supervision does not target transition onset directly, and evaluation does not measure boundary localization accuracy. The present work differs at the task, annotation, and metric levels rather than solely at the model, training, or sensor-fusion level.

\subsection{Changepoint Detection and Temporal Segmentation}
Classical temporal segmentation methods, including Pruned Exact Linear Time (PELT), Bayesian changepoint detection, and Hidden semi-Markov models (HSMMs) with explicit duration modeling~\cite{killick2012optimal,adams2007bayesian,yu2010hidden}, provide natural tools for boundary localization in structured time series.
They have been used in general segmentation and physiological activity recognition~\cite{bao2004activity}, but not systematically for sleep stage boundary localization from PPG under coarse 30-second supervision, where the goal is not only to mark a change but also to infer typed stage transitions with plausible dwell times.
Our label expansion pipeline adapts HSMM inference and local changepoint refinement to this setting while enforcing physiological transition constraints.

\subsection{Task Definition and Evaluation Protocols for Physiological Time Series}

In several physiological time-series domains, the shift from window-level classification to temporal localization has produced more informative tasks and benchmarks~\cite{bulling2014tutorial,neverova2015moddrop,hannun2019cardiologist,carter2024sleepvst}.
For example, event-level formulations in arrhythmia analysis and sleep-event assessment make timing errors directly measurable rather than averaging them over fixed windows~\cite{hannun2019cardiologist,carter2024sleepvst}.
Sleep analysis, however, remains centered on epoch-level classification~\cite{berry2017aasm}, and its standard metrics are insensitive to boundary timing.
SSTD is intended to address this gap for PPG sleep analysis by defining a transition-detection task together with metrics that measure temporal precision and event-level completeness.

\begin{figure}[t]
  \centering
  \includegraphics[width=\linewidth]{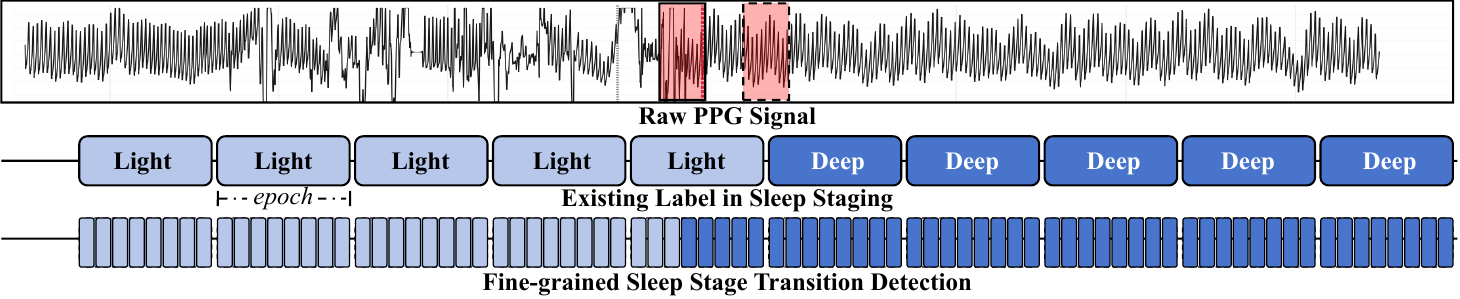}
\caption{Illustration of the annotation gap addressed by this work.
\textbf{Top:} a raw PPG recording containing a Light$\to$Deep transition; the highlighted region marks the signal perturbation at the boundary.
\textbf{Middle (Existing Label):} conventional sleep staging assigns a single coarse label per 30-second epoch, placing the transition arbitrarily at the epoch boundary with no sec-level precision.
\textbf{Bottom (SSTD):} our proposed sec-level labels, produced by the label expansion pipeline, explicitly locate the transition onset within the epoch and assign per-second stage assignments throughout.}
  \label{fig:label_compare} 
\end{figure}

\section{Sleep Stage Transition Detection: Evidence, Task, and Dataset}\label{sec:sstd}


\subsection{Evidence for PPG's Transition-Sensitive Characteristics}\label{subsec:evidence}

\textbf{Physiological basis.}
Section~\ref{sec:intro} identified a characteristic asymmetry in PPG's sleep-related information content.
Two cardiovascular mechanisms underlie this property: the inter-beat interval sequence reflects beat-to-beat sympathovagal shifts on second-to-second timescales~\cite{malik1996heart}, and pulse waveform morphology captures concurrent changes in peripheral vascular resistance with similarly rapid onset~\cite{allen2007photoplethysmography}.
By contrast, EEG spectral patterns accumulate over tens of seconds~\cite{werth1997spindle}.

\begin{figure}[t]
  \centering
  \includegraphics[width=\linewidth]{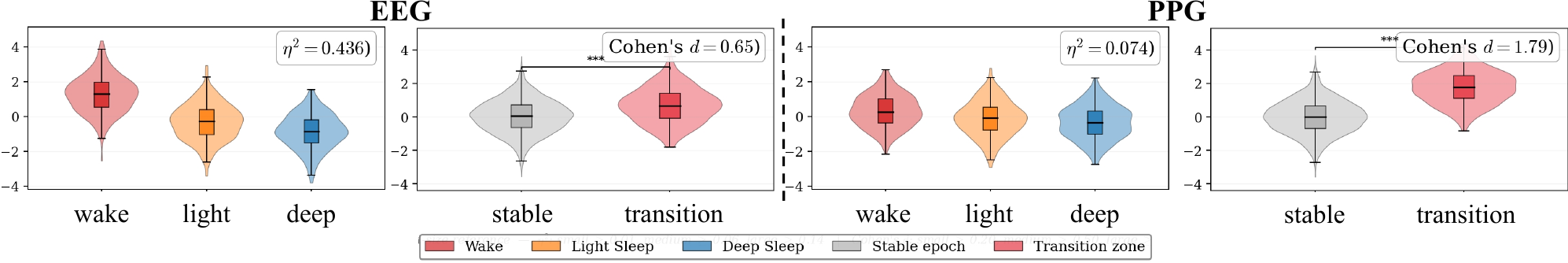}
\caption{Population-level feature distributions on separate MESA subjects for EEG (left) and PPG (right).
Each modality is summarized by inter-stage separability across stable epochs ($\eta^2$) and boundary sensitivity between stable epochs and $\pm5$\,s transition zones (Cohen's $d$).
EEG yields high stage separability but weaker boundary response, whereas PPG shows lower stage separability but stronger boundary sensitivity.}
  \label{fig:signal_compare} 
\end{figure}

\begin{table}[h]
\centering
\caption{Cross-modal transition sensitivity on 1000 transition events from 100 held-out MESA subjects.}
\label{tab:sensitivity}
\begin{adjustbox}{width=0.6\textwidth,center}
\setlength{\tabcolsep}{8pt}
\begin{tabular}{lccc}
\toprule
\textbf{Modality} & \textbf{Var.\ ratio} $\uparrow$ & \textbf{TDL (s)} $\downarrow$ & \textbf{BDI} $\uparrow$ \\
\midrule
PPG (HRV + morph.)   & \textbf{1.84} & \textbf{2.3} & \textbf{1.62} \\
EEG      & 1.31 & 5.1 & 1.18 \\ 
\bottomrule
\end{tabular}
\end{adjustbox}
\end{table}
\textbf{Cross-modal comparison at transition boundaries.}
To quantify this asymmetry, we conducted a controlled comparison using concurrent overnight PPG and EEG recordings from MESA~\cite{chen2015racial}.
For each annotated transition, we compared feature variability in a $\pm$5\,s boundary window against a matched 10-second stable-epoch reference window.
The resulting normalized variance ratio is above 1 when boundary windows are more variable than stable windows.
Figure~\ref{fig:signal_compare} shows the corresponding transition-versus-stable separation on held-out MESA subjects.
Quantitative analysis across 1,000 transition events from 100 held-out MESA subjects yields $\bar{r}_\text{PPG} = 1.84$, $\bar{r}_\text{EEG} = 1.31$, indicating that, under the feature representations examined here, \textit{PPG features exhibit a higher boundary-to-stable contrast ratio than EEG features}.
As a modality-normalized statistic, the variance ratio should be interpreted as relative boundary prominence rather than absolute signal richness.
Under this measure, abrupt PPG perturbations stand out more clearly against its quieter stable-epoch baseline, whereas EEG combines richer stable-epoch structure with more gradual transitions~\cite{trinder2001autonomic,malik1996heart,werth1997spindle}.
Table~\ref{tab:sensitivity} reports two complementary metrics: Transition Detection Latency (TDL) measures how quickly features respond after a reference transition, and the Boundary Discriminability Index (BDI) summarizes separation between boundary and matched stable windows; formal definitions are given in Appendix~\ref{app:metrics}.

\textbf{Implications.}
Two complementary findings summarize the characteristic asymmetry relevant to task design.
\textbf{(F1)~Boundary sensitivity:} under the HRV and morphology features examined here, PPG features exhibit a \textbf{higher boundary-to-stable contrast ratio} than EEG, with $\bar{r}_\text{PPG} = 1.84$ compared to $1.31$ for EEG signals.
\textbf{(F2)~Epoch weakness:} within stable epochs, PPG's inter-stage separability is \textbf{lower than EEG's}, with $\eta^2 = 0.07$ compared to $0.44$ for EEG as shown in Figure~\ref{fig:signal_compare}.
Together, F1 and F2 indicate that PPG's informational advantage, under the feature representations examined here, is concentrated at transition boundaries, which coincide with the temporal locations that 30-second epoch classification averages over.
A task formulation that explicitly targets transition events is therefore a plausible way to align the benchmark with this intrinsic sensitivity.

\subsection{Task Formalization}\label{subsec:task}

\textbf{Definition.}
Let $X = \{x_t\}_{t=1}^{T}$ denote a continuous PPG recording at sec-level sampling resolution.
Sleep Stage Transition Detection (SSTD) is defined as the task of predicting, from $X$, the set of transition events $\mathcal{E} = \{(\tau_k,\, s_k^-,\, s_k^+)\}_{k=1}^{K}$, where $\tau_k \in \{1, \ldots, T\}$ is the onset time of the $k$-th transition in seconds, $s_k^-$ is the stage immediately before the transition, and $s_k^+ \neq s_k^-$ is the stage immediately after.
Stage labels are drawn from a four-class vocabulary $\mathcal{S} = \{\text{Wake},\, \text{Light},\, \text{Deep},\, \text{REM}\}$.

\textbf{Four-class stage vocabulary.}
The standard AASM taxonomy is partially collapsed: Wake is retained; N1 and N2 are merged into \textit{Light} sleep; N3 is retained as \textit{Deep} sleep; and REM is retained explicitly.
This design preserves the clinically important REM state and its associated transition patterns while removing the notoriously unreliable N1/N2 boundary, and therefore serves as the main benchmark choice for the present single-modality PPG setting~\cite{berry2017aasm,fonseca2015sleep}.
Appendix~\ref{app:extended_vocab} reports a five-class extension that separates N1 and N2, showing that the same framework remains applicable under the full AASM taxonomy but with lower label quality and weaker downstream performance.

\textbf{Relationship to conventional staging.}
SSTD and epoch-level staging are \textit{complementary}: a complete SSTD prediction implies a piecewise-constant stage assignment collapsible to epoch labels by majority vote, while consecutive differing epoch labels identify candidate windows for SSTD refinement.

\textbf{Evaluation metrics.}
We introduce two dedicated metrics that quantify aspects of prediction quality:

\textit{Boundary Error (BE).}
For each predicted transition event $\hat\tau$ matched to a reference event $\tau^*$ of the same type within a matching tolerance $\delta_\text{match}$, the boundary error is $\text{BE} = |\hat\tau - \tau^*|$ in seconds.
We report the mean BE and its standard deviation over all matched pairs.
Unmatched predicted events and unmatched reference events contribute to false positives and false negatives respectively.

\textit{Transition F1 (TF1).}
A predicted event is classified as a true positive if it falls within tolerance $\delta_\text{match}$ of a reference event of matching transition type; otherwise it is a false positive.
Reference events with no matching prediction within tolerance are false negatives.
Transition precision, recall, and F1 are computed from these counts, so TF1 is an event-level F1 score over typed transitions.
Together, TF1 measures detection completeness and type accuracy while BE measures temporal localization sharpness, two dimensions that standard epoch-level metrics cannot capture.
In the present benchmark, the sec-level reference events are produced by the label expansion pipeline described in Section~\ref{subsec:dataset} and externally checked against expert review before downstream use.
We report TF1 at $\delta_\text{match} = 10$\,s in the main text and provide a sensitivity study over $\delta_\text{match}$ in Appendix~\ref{app:tolerance}.

\subsection{Label Expansion: Constructing Sec-Level SSTD Annotations}\label{subsec:dataset}

\textbf{Overview.}
We develop a \textit{two-stage label expansion pipeline} that converts the coarse 30-second epoch labels available in existing datasets into sec-level pseudo-labels.

\textbf{Stage 1: Hidden semi-Markov model.}
Sleep stages persist for several minutes, yet a standard HMM permits a state change at every time step.
An HSMM addresses this by introducing an explicit duration distribution $P(d \mid s)$ for each stage $s$, constraining dwell times to physiologically plausible ranges~\cite{yu2010hidden}.
We define an HSMM over PPG feature vectors emitted at 1-second stride\footnote{The feature sequence is indexed at 1-second resolution, but HRV-derived quantities that require longer temporal support are computed from rolling 30-second buffers updated every second rather than from isolated 1-second segments. Frequency-domain HRV is therefore defined over buffered context rather than single-second windows. The full feature specification is provided in the appendix.} with Gaussian emissions and Poisson duration distributions; the transition matrix is estimated from the observed 30-second label sequences, and coarse epoch-level sleep stage labels enter as multiplicative soft constraints: for each second $t$, the emission likelihood $P(x_t \mid s_t)$ is scaled by $\lambda$ when $s_t$ agrees with the enclosing 30-second epoch label $\ell_t$ and by $1/\lambda$ otherwise, with $\lambda = 3$; sensitivity to this parameter is reported in Appendix~\ref{app:label_ablation}.
The joint inference objective is:
\begin{equation}
\small
P(s_{1:T},\, d_{1:K} \mid X, L) \;=\; \prod_{k=1}^{K} P(d_k \mid s_k) \cdot \prod_{t=1}^{T} \bigl[P(x_t \mid s_t)\cdot c(s_t,\ell_t)\bigr] \cdot P(s_k \mid s_{k-1}),
\label{eq:hsmm}
\end{equation}
where $d_k$ is the dwell time of segment $k$, $c(s_t,\ell_t)=\lambda$ if $s_t=\ell_t$ and $1/\lambda$ otherwise encodes the soft coarse-label constraint, $L=\{\ell_t\}$ denotes the coarse epoch labels, and $K$ is the number of contiguous state segments.
Viterbi decoding produces an initial sec-level state sequence with physiologically plausible dwell-time structure.

\textbf{Stage 2: Local changepoint refinement.}
The HSMM provides sec-level state sequences, but its smoothed emission model can blur the exact onset timing that SSTD ultimately evaluates.
For each HSMM-identified boundary $\tau$, we search within a local window $[\tau - \Delta, \tau + \Delta]$ for the refined position $\tau^*$ that maximizes the evidence of a distributional shift:
\begin{equation}
\small
\tau^* = \arg\max_{\tau' \in [\tau-\Delta,\,\tau+\Delta]} \;\left\|\mu_{\text{before}}(\tau') - \mu_{\text{after}}(\tau')\right\| \cdot \left(1 + \beta \cdot \frac{\sigma_{\text{after}}(\tau')}{\sigma_{\text{before}}(\tau')}\right),
\label{eq:bcd}
\end{equation}
where $\mu_{\text{before/after}}$ and $\sigma_{\text{before/after}}$ are the mean and standard deviation of feature vectors in 10-second windows on either side of $\tau'$, and $\beta$ weights the post-transition variance increase.
We set $\Delta = 15$\,s and $\beta = 0.5$; sensitivity of the expanded labels to these parameters is analyzed in Appendix~\ref{app:label_ablation}.

\textbf{Physiological transition constraints.}
In normal adult sleep, direct Wake$\leftrightarrow$Deep transitions do not occur without traversing an intermediate Light-sleep stage~\cite{carskadon2005normal}.
A post-processing step enforces this constraint by inserting a brief Light-sleep segment at any detected Wake-to-Deep or Deep-to-Wake boundary, with the inserted duration set to the minimum credible Light-stage dwell time estimated from the training population.

\textbf{Label quality assessment.}
We evaluate expanded labels using three metrics: (1) \textit{window consistency}, the fraction of 30-second windows in which the majority pseudo-label agrees with the corresponding PSG epoch label; (2) \textit{implausible transition rate}, the fraction of consecutive state pairs that violate physiological progression rules; and (3) \textit{pseudo-transition density}, the number of adjacent stage changes per hour in the sec-level pseudo-label sequence.
The third quantity is used as a fragmentation diagnostic rather than as a direct estimate of clinical macro-stage dwell times.

Quantitative assessment of label quality on the evaluation cohorts is reported in Section~\ref{sec:experiments}.

\textbf{Applicability.}
The pipeline is designed for PSG-scored datasets that provide epoch-level labels and concurrent PPG recordings, without requiring dataset-specific architectural modifications; cohort-specific instantiation details are provided in Section~\ref{sec:experiments}.

\section{Experiments}\label{sec:experiments}


\subsection{Experimental Setup}\label{subsec:setup}

\textbf{Datasets.}
All training and primary evaluation use the MESA (Multi-Ethnic Study of Atherosclerosis)~\cite{chen2015racial} cohort: 2,108 community-dwelling adults aged 54--93 and broad ethnic diversity of 38\% White, 28\% Black, 23\% Hispanic, and 11\% Asian; PPG is sampled at 64\,Hz from finger pulse oximetry.
The label expansion pipeline of Section~\ref{subsec:dataset} is applied to MESA's AASM-scored PSG epoch labels to produce sec-level SSTD annotations under the four-class vocabulary.
MESA provides the original 30-second PSG labels used for the primary conventional sleep-staging comparison after majority-vote collapse of per-second predictions.
As supplementary cross-cohort validation, we evaluate on the Cleveland Family Study (CFS)~\cite{redline1995familial}: 730 subjects aged 6--88 scored under the Rechtschaffen \& Kales protocol rather than AASM.
CFS introduces two compounding sources of domain shift, population differences and annotation-protocol differences, and is reported separately in Appendix~\ref{app:cross_dataset}.
All models evaluated on CFS are trained exclusively on MESA and applied zero-shot.

\textbf{Pipeline validation dataset.}
To assess label fidelity before large-scale label expansion, we use a separate expert-reviewed PPG dataset with concurrent PSG and sec-level sleep stage boundaries.
This dataset serves to validate the label expansion procedure before it is applied to MESA and CFS; it plays no role in model training, model selection, or baseline comparison.
We report the main agreement statistics in the body text, while the full review protocol and adjudication procedure are provided in Appendix~\ref{app:clinical_validation}.

\textbf{Evaluation metrics.}
\textit{SSTD metrics}: Boundary Error (BE, seconds $\downarrow$) and Transition F1 (TF1 $\uparrow$) with $\delta_\text{match} = 10$\,s (Section~\ref{subsec:task}).
\textit{Epoch metrics}: Accuracy (Acc) and Macro-F1 (MF1) for four-class stage prediction via majority-vote collapse to 30-second epochs.
Unless otherwise stated, conventional staging comparisons are reported against the original PSG-scored epoch labels after this collapse, whereas BE and TF1 are used to analyze transition localization under the expanded-label setting; Appendix~\ref{app:tolerance} reports sensitivity to $\delta_\text{match}$.

\textbf{Baseline methods.}
We compare LightSleep and five representative existing approaches spanning four design paradigms (Table~\ref{tab:main_results}).
\textit{Simple reference baseline}: \textbf{LightSleep}, a deliberately lightweight 1D CNN comprising four temporal convolution blocks with max pooling and a linear classification head. It is intentionally kept simple, with no attention mechanism, multi-branch decomposition, or hand-crafted task-specific inductive bias, so that the contribution of sec-level supervision can be assessed without relying on an unusually strong architecture.
\textit{Feature-based}: \textbf{HRV-RF}~\cite{fonseca2023computationally}, a random forest operating on hand-crafted HRV statistics per epoch.
\textit{End-to-end waveform}: \textbf{SleepPPGNet}~\cite{kotzen2022sleepppg}, a CNN--RNN architecture trained on raw PPG; \textbf{PPG-TCN}~\cite{huttunen2021assessment}, a temporal convolutional network with dilated causal convolutions.
\textit{Time-frequency}: \textbf{SleepPPGNet2}~\cite{attia2025sleepppg}, a model that employs multi-source domain training strategies; \textbf{CWT-ResNet}, a ResNet-18 encoder applied to CWT scalograms as a strong time-frequency reference.
These baselines provide representative coverage of major single-modality PPG families. All deep baselines are retrained as matched variants under the same input segmentation and optimization setup. We focus on methods that admit matched epoch-only and SSTD supervision so that differences can be attributed to the supervision as cleanly as possible.

\subsection{Label Expansion Validation}\label{subsec:label_validation}

The practical role of this label expansion pipeline in this paper is to provide sec-level supervision for downstream training on public datasets.
Before using it in that role, we validate label quality from three complementary angles that each avoid circularity.

\textbf{Automatic consistency metrics.}
Table~\ref{tab:label_quality} reports two automatic quality metrics computed against the original PSG-scored epoch labels, which are independent of the expansion pipeline.
On MESA, the full pipeline achieves window consistency of 0.857 and reduces the implausible transition rate to 7.83\%, compared to 41.4\% for the HSMM-only Stage~1 baseline; the N1/N2 merge into a single Light class and the physiological-constraint post-processing jointly account for this improvement~\cite{danker2009interrater,carskadon2005normal}.
On CFS, window consistency remains at 0.839 despite the R\&K-to-AASM annotation-protocol shift, suggesting that the pipeline retains similar coarse-label agreement under a dataset with different labeling rules.
Window consistency measures 30-second coarse agreement and functions as a lower bound on label fidelity; it does not directly assess sec-level boundary placement.
Pseudo-transition density is likewise interpreted comparatively rather than normatively: it counts every adjacent state change in the sec-level pseudo-label sequence, including brief intermediary segments such as the Light stage inserted to eliminate physiologically impossible Wake$\leftrightarrow$Deep jumps. We therefore read this quantity jointly with the implausible transition rate and the external expert agreement reported below, rather than as a standalone estimate of macro-stage dwell time.

\begin{table}[h]
\centering
\caption{Label expansion quality. Pipeline ablation on 100 separate MESA subjects (full ablation in Appendix~\ref{app:label_ablation}) and cross-cohort evaluation on CFS (R\&K-scored).}
\label{tab:label_quality}
\begin{adjustbox}{width=0.8\textwidth,center}
\setlength{\tabcolsep}{6pt}
\begin{tabular}{lccc}
\toprule
\textbf{Configuration} & \textbf{Win. consist.} $\uparrow$ & \textbf{Implaus. rate} $\downarrow$ & \textbf{Pseudo-trans./hr} \\
\midrule
HSMM only (5-class, MESA) & 0.825 & 41.4\% & 8.31 \\
Full pipeline (MESA)       & \textbf{0.857} & \textbf{7.83\%} & 7.66 \\
\midrule
Full pipeline (CFS)        & 0.839 & 9.24\% & 8.17 \\
\bottomrule
\end{tabular}
\end{adjustbox}
\end{table}


The most direct external check compares pipeline outputs on the separate expert-reviewed dataset against an independently established expert reference.
Using the same hyperparameters as on MESA, the pipeline reaches $\kappa = 0.81$ and TF1 = 78.6\% against expert consensus, with a mean boundary error of 4.2\,s (Table~\ref{tab:expert_agreement}).
These statistics support the clinical plausibility of the expanded labels and provide an external check that is fully separate from the downstream MESA and CFS experiments.
Review protocol, adjudication procedure, and additional agreement analysis are provided in Appendix~\ref{app:clinical_validation}.


\begin{table}[h]
\centering
\caption{Expert annotation agreement on the separate validation set. }
\label{tab:expert_agreement}
\begin{adjustbox}{width=0.5\textwidth,center}
\setlength{\tabcolsep}{6pt}
\begin{tabular}{lcc}
\toprule
\textbf{Comparison} & $\boldsymbol{\kappa}$ & \textbf{TF1 (\%)} \\
\midrule
Pipeline vs.\ Expert consensus & 0.81 & 78.6 \\
Expert inter-rater agreement   & 0.89 & --- \\
\bottomrule
\end{tabular}
\end{adjustbox}
\end{table}

\textbf{Sleep-wake detection as a functional label-quality check.}
A second non-circular check uses sleep-wake detection on MESA, a binary task with established evaluation criteria whose ground-truth labels are the original PSG wake/sleep annotations, independent of the expansion pipeline.
If the expanded sec-level labels encode genuine physiological boundary information rather than pipeline artefacts, features derived from those labels should provide a training signal that transfers to this independent binary task.
We evaluate this transfer on LightSleep and on SleepPPGNet as an existing baseline; results and analysis are in Table~\ref{tab:sleep_wake}.
We use sleep-wake detection here only as an auxiliary non-circular validation task. The aim is not to claim a new state-of-the-art solution for wearable sleep-wake monitoring, but to test whether the expanded sec-level labels transfer useful information to an independently supervised task whose labels are available.

\begin{table}[h]
\centering
\caption{Sleep-wake detection results on MESA (Wake vs.\ Sleep binary classification). Epoch-only: trained on 30-second epoch labels only. SSTD$\to$Sleep Wake: SSTD pre-trained then fine-tuned on sleep-wake labels. Metrics are against original epoch labels (independent of the expansion pipeline).}
\label{tab:sleep_wake}
\begin{adjustbox}{width=0.6\textwidth,center}
\setlength{\tabcolsep}{8pt}
\begin{tabular}{lcc}
\toprule
\textbf{Method} & \textbf{Acc (\%)} & \textbf{F1 (\%)} \\
\midrule
Epoch-only (SleepPPGNet) & \tempresult{87.3} & \tempresult{83.6} \\
SSTD$\to$Sleep-Wake (SleepPPGNet) & \tempresult{92.6} & \tempresult{89.1} \\
\midrule
Epoch-only (LightSleep) & 85.1 & 80.8 \\
SSTD$\to$Sleep-Wake (LightSleep) & 89.3 & 85.3 \\
\bottomrule
\end{tabular}
\end{adjustbox}
\end{table}

\subsection{Transition-Level Analysis on MESA}\label{subsec:comparison}

\begin{table}[h]
\centering
\caption{Transition-level comparison on the MESA four-class setting.
LightSleep is trained against the expanded SSTD labels first, then trained and evaluated with other models against MESA's original 30s labels; this analysis examines whether the learned representations capture boundary timing.}
\label{tab:main_results}
\begin{adjustbox}{width=0.8\textwidth,center}
\setlength{\tabcolsep}{10pt}
\begin{tabular}{lcccc}
\toprule
\textbf{Method} & \textbf{BE (s)} $\downarrow$ & \textbf{TF1 (\%)} $\uparrow$ & \textbf{Acc (\%)} $\uparrow$ & \textbf{MF1 (\%)} $\uparrow$ \\
\midrule
HRV-RF~\cite{fonseca2023computationally}              & 9.7 & 38.2 & 68.4 & 54.1 \\
SleepPPGNet~\cite{kotzen2022sleepppg}          & 7.4 & 46.5 & 74.1 & 62.3 \\
PPG-TCN~\cite{huttunen2021assessment}              & 6.8 & 49.1 & 73.6 & 63.8 \\
SleepPPGNet2~\cite{attia2025sleepppg} & 6.5 & 50.3 & 74.8 & 64.2 \\
CWT-ResNet           & 6.1 & 52.7 & 71.3 & 62.0 \\
\midrule
LightSleep        & \tempresult{\textbf{4.6}} & \tempresult{\textbf{59.4}} & \tempresult{\textbf{77.2}} & \tempresult{\textbf{68.5}} \\
\bottomrule
\end{tabular}
\end{adjustbox}
\end{table}

Table~\ref{tab:main_results} provides a transition-level analysis under the expanded-label setting; the primary conventional sleep-staging comparison against MESA's original PSG epoch labels is deferred to Table~\ref{tab:transfer}. Two patterns emerge.
First, SSTD metrics differentiate methods that epoch metrics cannot: SleepPPGNet and PPG-TCN achieve similar epoch accuracy (74.1\% vs.\ 73.6\%) yet differ in TF1 (46.5\% vs.\ 49.1\%) and BE (7.4\,s vs.\ 6.8\,s), confirming the diagnostic value of transition-level evaluation.
Second, lower boundary error is generally accompanied by higher epoch accuracy, although the ordering is not perfectly monotonic across architectures. Notably, even the deliberately lightweight LightSleep baseline attains the strongest results in this comparison (BE \tempresult{4.6}\,s, TF1 \tempresult{59.4}\%, Acc \tempresult{77.2}\%, MF1 \tempresult{68.5}\%), suggesting that the value of sec-level supervision is not confined to unusually high-capacity or heavily engineered models.
The sensitivity analysis to matching tolerance $\delta_\text{match}$ in Appendix~\ref{app:tolerance} confirms that these rankings are preserved across $\delta \in \{5, 10, 15\}$\,s.

\subsection{Transfer to Epoch-Level Staging}\label{subsec:ablation}
\begin{table}[h]
\caption{Transfer benefit of SSTD pre-training on conventional 30\,s four-class epoch staging.}
\label{tab:transfer}
\centering
\footnotesize
\setlength{\tabcolsep}{10pt}
\begin{adjustbox}{width=0.7\textwidth,center}
\begin{tabular}{lcccc}
\toprule
\multirow{2}{*}{\textbf{Method}} & \multicolumn{2}{c}{\textbf{Epoch-only}} & \multicolumn{2}{c}{\textbf{SSTD$\to$Epoch}} \\
\cmidrule(lr){2-3} \cmidrule(lr){4-5}
& Acc & MF1 & Acc\,[$\Delta$] & MF1\,[$\Delta$] \\
\midrule
HRV-RF   & 68.4 & 54.1 & 73.5\,[+5.1] & 57.0\,[+2.9] \\
SleepPPGNet   & 74.1 & 62.3 & 79.8\,[+5.7] & 67.0\,[+4.7] \\
PPG-TCN       & 73.6 & 63.8 & 78.3\,[+4.7] & 65.5\,[+1.7] \\
SleepPPGNet2       & 74.8 & 64.2 & \textbf{80.3}\,[+5.5] & \textbf{68.5}\,[+4.3] \\
CWT-ResNet    & 71.3 & 62.0 & 75.0\,[+3.7] & 64.9\,[+2.9] \\
LightSleep & \tempresult{72.4} & \tempresult{62.3} & \tempresult{\tempresult{77.2}}\,[\tempresult{+4.8}] & \tempresult{\tempresult{68.5}}\,[\tempresult{+6.2}] \\
\bottomrule
\end{tabular}
\end{adjustbox}
\end{table}

\textbf{Transfer (Table~\ref{tab:transfer}).}
Table~\ref{tab:transfer} is the primary practical comparison for conventional sleep staging because it evaluates models against MESA's original PSG epoch labels after collapsing per-second predictions.
Within each matched architecture pair, SSTD pre-training consistently improves four-class epoch-level staging across all six evaluated architectures.
This transfer is practically meaningful because the hardest 30-second windows cluster near true stage transitions, where the majority label is most sensitive to boundary placement. A representation that resolves second-level boundary structure more accurately therefore provides cleaner evidence for conventional epoch prediction.
Crucially, the epoch-level gains are evaluated against the original PSG epoch labels, which are independent of the pseudo-label pipeline, thereby decoupling the transfer benefit from any label-construction choices and mitigating concerns about circularity between the expansion process and downstream evaluation.
Appendix~\ref{app:cross_dataset} extends this analysis to CFS under a zero-shot cross-cohort protocol, showing that the transfer benefit is preserved despite compounding demographic and annotation-protocol shifts.

\section{Conclusion}\label{sec:conclusion}

This work introduces Sleep Stage Transition Detection (SSTD), motivated by a characteristic asymmetry in PPG: stage-specific differences are comparatively weak within stable epochs, yet transition-related perturbations are sharper on sec-level timescales than in the EEG reference examined here.
A label expansion pipeline converts coarse PSG-scored epoch labels into sec-level SSTD annotations; dedicated BE and TF1 metrics expose performance differences invisible to epoch-level evaluation.
Even with a deliberately lightweight 1D CNN baseline, LightSleep, sec-level supervision remains effective: LightSleep attains \tempresult{4.6}\,s BE and \tempresult{59.4}\% TF1 on the four-class MESA benchmark, outperforming the more complex representative baselines included in the current evaluation protocol.
SSTD pre-training improves conventional four-class epoch-level staging by 1.7--4.7\,pp in MF1 across five representative existing baselines against MESA's original PSG epoch labels, with the transfer benefit preserved under supplementary zero-shot cross-cohort evaluation on CFS (Appendix~\ref{app:cross_dataset}).
Limitations of the current work are discussed in Appendix~\ref{app:limitations}.

\bibliography{example_paper}

@article{radha2019sleep,
  title={Sleep stage classification from heart-rate variability using long short-term memory neural networks},
  author={Radha, Mustafa and Fonseca, Pedro and Moreau, Arnaud and Ross, Marco and Cerny, Andreas and Anderer, Peter and Long, Xi and Aarts, Ronald M},
  journal={Scientific reports},
  volume={9},
  number={1},
  pages={14149},
  year={2019},
  publisher={Nature Publishing Group UK London}
}

@article{fonseca2015sleep,
  title={Sleep stage classification with ECG and respiratory effort},
  author={Fonseca, Pedro and Long, Xi and Radha, Mustafa and Haakma, Reinder and Aarts, Ronald M and Rolink, J{\'e}r{\^o}me},
  journal={Physiological measurement},
  volume={36},
  number={10},
  pages={2027--2040},
  year={2015},
  publisher={IOP Publishing}
}

@article{carskadon2005normal,
  title={Normal human sleep: an overview},
  author={Carskadon, Mary A and Dement, William C and others},
  journal={Principles and practice of sleep medicine},
  volume={4},
  number={1},
  pages={13--23},
  year={2005},
  publisher={Elsevier Saunders Philadelphia, PA}
}

@article{werth1997spindle,
  title={Spindle frequency activity in the sleep EEG: individual differences and topographical distribution},
  author={Werth, Esther and Achermann, Peter and Dijk, Derk-Jan and Borb{\'e}ly, Alexander A},
  journal={Electroencephalography and clinical neurophysiology},
  volume={103},
  number={5},
  pages={535--542},
  year={1997},
  publisher={Elsevier}
}

@article{allen2007photoplethysmography,
  title={Photoplethysmography and its application in clinical physiological measurement},
  author={Allen, John},
  journal={Physiological measurement},
  volume={28},
  number={3},
  pages={R1--R39},
  year={2007}
}

@article{malik1996heart,
  title={Heart rate variability: Standards of measurement, physiological interpretation, and clinical use: Task force of the European Society of Cardiology and the North American Society for Pacing and Electrophysiology},
  author={Malik, Marek},
  journal={Annals of Noninvasive Electrocardiology},
  volume={1},
  number={2},
  pages={151--181},
  year={1996},
  publisher={Wiley Online Library}
}

@misc{berry2017aasm,
  title={AASM scoring manual updates for 2017 (version 2.4)},
  author={Berry, Richard B and Brooks, Rita and Gamaldo, Charlene and Harding, Susan M and Lloyd, Robin M and Quan, Stuart F and Troester, Matthew T and Vaughn, Bradley V},
  journal={Journal of clinical sleep medicine},
  volume={13},
  number={5},
  pages={665--666},
  year={2017},
  publisher={American Academy of Sleep Medicine}
}

@article{hannun2019cardiologist,
  title={Cardiologist-level arrhythmia detection and classification in ambulatory electrocardiograms using a deep neural network},
  author={Hannun, Awni Y and Rajpurkar, Pranav and Haghpanahi, Masoumeh and Tison, Geoffrey H and Bourn, Codie and Turakhia, Mintu P and Ng, Andrew Y},
  journal={Nature medicine},
  volume={25},
  number={1},
  pages={65--69},
  year={2019},
  publisher={Nature Publishing Group US New York}
}

@article{neverova2015moddrop,
  title={Moddrop: adaptive multi-modal gesture recognition},
  author={Neverova, Natalia and Wolf, Christian and Taylor, Graham and Nebout, Florian},
  journal={IEEE Transactions on Pattern Analysis and Machine Intelligence},
  volume={38},
  number={8},
  pages={1692--1706},
  year={2015},
  publisher={IEEE}
}

@article{bulling2014tutorial,
  title={A tutorial on human activity recognition using body-worn inertial sensors},
  author={Bulling, Andreas and Blanke, Ulf and Schiele, Bernt},
  journal={ACM Computing Surveys (CSUR)},
  volume={46},
  number={3},
  pages={1--33},
  year={2014},
  publisher={ACM New York, NY, USA}
}

@inproceedings{bao2004activity,
  title={Activity recognition from user-annotated acceleration data},
  author={Bao, Ling and Intille, Stephen S},
  booktitle={International conference on pervasive computing},
  pages={1--17},
  year={2004},
  organization={Springer}
}

@article{yu2010hidden,
  title={Hidden semi-Markov models},
  author={Yu, Shun-Zheng},
  journal={Artificial intelligence},
  volume={174},
  number={2},
  pages={215--243},
  year={2010},
  publisher={Elsevier}
}

@article{adams2007bayesian,
  title={Bayesian online changepoint detection},
  author={Adams, Ryan Prescott and MacKay, David JC},
  journal={arXiv preprint arXiv:0710.3742},
  year={2007}
}

@inproceedings{carter2024sleepvst,
  title={SleepVST: Sleep staging from near-infrared video signals using pre-trained transformers},
  author={Carter, Jonathan F and Jorge, Jo{\~a}o and Gibson, Oliver and Tarassenko, Lionel},
  booktitle={Proceedings of the IEEE/CVF Conference on Computer Vision and Pattern Recognition},
  pages={12479--12489},
  year={2024}
}

@article{morokuma2023deep,
  title={Deep learning-based sleep stage classification with cardiorespiratory and body movement activities in individuals with suspected sleep disorders},
  author={Morokuma, Seiichi and Hayashi, Toshinari and Kanegae, Masatomo and Mizukami, Yoshihiko and Asano, Shinji and Kimura, Ichiro and Tateizumi, Yuji and Ueno, Hitoshi and Ikeda, Subaru and Niizeki, Kyuichi},
  journal={Scientific reports},
  volume={13},
  number={1},
  pages={17730},
  year={2023},
  publisher={Nature Publishing Group UK London}
}

@article{quino2024optimizing,
  title={Optimizing Photoplethysmography-Based Sleep Staging Models by Leveraging Temporal Context for Wearable Devices Applications},
  author={Quino, Joseph AP and Cardenas, Diego AC and Toledo, Marcelo AF and Dias, Felipe M and Ribeiro, Estela and Krieger, Jos{\'e} E and Gutierrez, Marco A},
  journal={arXiv preprint arXiv:2410.00693},
  year={2024}
}

@article{markov2025interpretable,
  title={Interpretable feature-based machine learning for automatic sleep detection using photoplethysmography},
  author={Markov, Karmen and Elgendi, Mohamed and Birrer, Vera and Menon, Carlo},
  journal={npj Biosensing},
  volume={2},
  number={1},
  pages={24},
  year={2025},
  publisher={Nature Publishing Group UK London}
}

@article{coon2025getting,
  title={Getting More from Less: Transfer Learning Improves Sleep Stage Decoding Accuracy in Peripheral Wearable Devices},
  author={Coon, William G and Luna, Diego and Panagrahi, Akshita and Reid, Matthew and Ogg, Mattson},
  journal={arXiv preprint arXiv:2506.00730},
  year={2025}
}

@article{zhao2021multi,
  title={A multi-class automatic sleep staging method based on photoplethysmography signals},
  author={Zhao, Xiangfa and Sun, Guobing},
  journal={Entropy},
  volume={23},
  number={1},
  pages={116},
  year={2021},
  publisher={MDPI}
}

@article{killick2012optimal,
  title={Optimal detection of changepoints with a linear computational cost},
  author={Killick, Rebecca and Fearnhead, Paul and Eckley, Idris A},
  journal={Journal of the American Statistical Association},
  volume={107},
  number={500},
  pages={1590--1598},
  year={2012},
  publisher={Taylor \& Francis}
}

@article{chambon2018deep,
  title={A deep learning architecture for temporal sleep stage classification using multivariate and multimodal time series},
  author={Chambon, Stanislas and Galtier, Mathieu N and Arnal, Pierrick J and Wainrib, Gilles and Gramfort, Alexandre},
  journal={IEEE Transactions on Neural Systems and Rehabilitation Engineering},
  volume={26},
  number={4},
  pages={758--769},
  year={2018},
  publisher={IEEE}
}

@article{supratak2017deepsleepnet,
  title={DeepSleepNet: A model for automatic sleep stage scoring based on raw single-channel EEG},
  author={Supratak, Akara and Dong, Hao and Wu, Chao and Guo, Yike},
  journal={IEEE transactions on neural systems and rehabilitation engineering},
  volume={25},
  number={11},
  pages={1998--2008},
  year={2017},
  publisher={IEEE}
}

@article{phan2023seqsleepnet,
  title={L-SeqSleepNet: Whole-cycle long sequence modeling for automatic sleep staging},
  author={Phan, Huy and Lorenzen, Kristian P and Heremans, Elisabeth and Ch{\'e}n, Oliver Y and Tran, Minh C and Koch, Philipp and Mertins, Alfred and Baumert, Mathias and Mikkelsen, Kaare B and De Vos, Maarten},
  journal={IEEE Journal of Biomedical and Health Informatics},
  volume={27},
  number={10},
  pages={4748--4757},
  year={2023},
  publisher={IEEE}
}

@article{sors2018convolutional,
  title={A convolutional neural network for sleep stage scoring from raw single-channel EEG},
  author={Sors, Arnaud and Bonnet, St{\'e}phane and Mirek, S{\'e}bastien and Vercueil, Laurent and Payen, Jean-Fran{\c{c}}ois},
  journal={Biomedical Signal Processing and Control},
  volume={42},
  pages={107--114},
  year={2018},
  publisher={Elsevier}
}

@article{kotzen2022sleepppg,
  title={SleepPPG-Net: A deep learning algorithm for robust sleep staging from continuous photoplethysmography},
  author={Kotzen, Kevin and Charlton, Peter H and Salabi, Sharon and Amar, Lea and Landesberg, Amir and Behar, Joachim A},
  journal={IEEE Journal of Biomedical and Health Informatics},
  volume={27},
  number={2},
  pages={924--932},
  year={2022},
  publisher={IEEE}
}

@article{fonseca2023computationally,
  title={A computationally efficient algorithm for wearable sleep staging in clinical populations},
  author={Fonseca, Pedro and Ross, Marco and Cerny, Andreas and Anderer, Peter and van Meulen, Fokke and Janssen, Hennie and Pijpers, Angelique and Dujardin, Sylvie and van Hirtum, Pauline and van Gilst, Merel and others},
  journal={Scientific Reports},
  volume={13},
  number={1},
  pages={9182},
  year={2023},
  publisher={Nature Publishing Group UK London}
}

@article{attia2024sleepppg,
  title={SleepPPG-Net2: Deep learning generalization for sleep staging from photoplethysmography},
  author={Attia, Shirel and Hershkovich, Revital Shani and Tabakhov, Alissa and Ang, Angeleene and Haimov, Sharon and Tauman, Riva and Behar, Joachim A},
  journal={arXiv preprint arXiv:2404.06869},
  year={2024}
}

@article{eldele2021attention,
  title={An attention-based deep learning approach for sleep stage classification with single-channel EEG},
  author={Eldele, Emadeldeen and Chen, Zhenghua and Liu, Chengyu and Wu, Min and Kwoh, Chee-Keong and Li, Xiaoli and Guan, Cuntai},
  journal={IEEE Transactions on Neural Systems and Rehabilitation Engineering},
  volume={29},
  pages={809--818},
  year={2021},
  publisher={IEEE}
}

@article{attia2025sleepppg,
  title={SleepPPG-Net2: Deep learning generalization for sleep staging from photoplethysmography},
  author={Attia, Shirel and Hershkovich, Revital Shani and Tabakhov, Alissa and Ang, Angeleene and Oksenberg, Arie and Tauman, Riva and Behar, Joachim A},
  journal={Physiological Measurement},
  volume={46},
  number={12},
  pages={125001},
  year={2025},
  publisher={IOP Publishing}
}

@article{habib2022performance,
  title={Performance of a convolutional neural network derived from PPG signal in classifying sleep stages},
  author={Habib, Ahsan and Motin, Mohammod Abdul and Penzel, Thomas and Palaniswami, Marimuthu and Yearwood, John and Karmakar, Chandan},
  journal={IEEE Transactions on Biomedical Engineering},
  volume={70},
  number={6},
  pages={1717--1728},
  year={2022},
  publisher={IEEE}
}

@article{korkalainen2020deep,
  title={Deep learning enables sleep staging from photoplethysmogram for patients with suspected sleep apnea},
  author={Korkalainen, Henri and Aakko, Juhani and Duce, Brett and Kainulainen, Samu and Leino, Akseli and Nikkonen, Sami and Afara, Isaac O and Myllymaa, Sami and T{\"o}yr{\"a}s, Juha and Lepp{\"a}nen, Timo},
  journal={Sleep},
  volume={43},
  number={11},
  pages={zsaa098},
  year={2020},
  publisher={Oxford University Press US}
}

@article{berry2012aasm,
  title={The AASM manual for the scoring of sleep and associated events},
  author={Berry, Richard B and Brooks, Rita and Gamaldo, Charlene E and Harding, Susan M and Marcus, Carole and Vaughn, Bradley V and others},
  journal={Rules, Terminology and Technical Specifications, Darien, Illinois, American Academy of Sleep Medicine},
  volume={176},
  number={2012},
  pages={7},
  year={2012}
}

@article{Zhai20,
author = {Zhai, Bing and Perez-Pozuelo, Ignacio and Clifton, Emma A. D. and Palotti, Joao and Guan, Yu},
title = {Making Sense of Sleep: Multimodal Sleep Stage Classification in a Large, Diverse Population Using Movement and Cardiac Sensing},
year = {2020},
issue_date = {June 2020},
publisher = {Association for Computing Machinery},
address = {New York, NY, USA},
volume = {4},
number = {2},
url = {https://doi.org/10.1145/3397325},
doi = {10.1145/3397325},
journal = {Proc. ACM Interact. Mob. Wearable Ubiquitous Technol.},
month = jun,
articleno = {67},
numpages = {33}
}

@article{sleepAI_review2020,
author = {Perez-Pozuelo, Ignacio and Zhai, Bing and Palotti, Joao and Mall, Raghvendra and Aupetit, Micha{\"e}l and Garcia-Gomez, Juan M. and Taheri, Shahrad and Guan, Yu and Fernandez-Luque, Luis},
date = {2020/03/23},
doi = {10.1038/s41746-020-0244-4},
id = {Perez-Pozuelo2020},
isbn = {2398-6352},
journal = {npj Digital Medicine},
number = {1},
pages = {42},
title = {The future of sleep health: a data-driven revolution in sleep science and medicine},
url = {https://doi.org/10.1038/s41746-020-0244-4},
volume = {3},
year = {2020}
}

@article{ting2005disorders,
  title={Disorders of sleep: an overview},
  author={Ting, Leon and Malhotra, Atul},
  journal={Primary care},
  volume={32},
  number={2},
  pages={305},
  year={2005}
}

@article{chen2015racial,
  title={Racial/ethnic differences in sleep disturbances: the Multi-Ethnic Study of Atherosclerosis (MESA)},
  author={Chen, Xiaoli and Wang, Rui and Zee, Phyllis and Lutsey, Pamela L and Javaheri, Sogol and Alc{\'a}ntara, Carmela and Jackson, Chandra L and Williams, Michelle A and Redline, Susan},
  journal={Sleep},
  volume={38},
  number={6},
  pages={877--888},
  year={2015},
  publisher={Oxford University Press}
}

@article{redline1995familial,
  title={The familial aggregation of obstructive sleep apnea.},
  author={Redline, Susan and Tishler, Peter V and Tosteson, Tor D and Williamson, John and Kump, Kenneth and Browner, Ilene and Ferrette, Veronica and Krejci, Patrick},
  journal={American journal of respiratory and critical care medicine},
  volume={151},
  number={3},
  pages={682--687},
  year={1995},
  publisher={American Public Health Association}
}

@article{sateia2014international,
  title={International classification of sleep disorders},
  author={Sateia, Michael J},
  journal={Chest},
  volume={146},
  number={5},
  pages={1387--1394},
  year={2014},
  publisher={Elsevier}
}

@article{somers1993sympathetic,
  title={Sympathetic-nerve activity during sleep in normal subjects},
  author={Somers, Virend K and others},
  journal={New England Journal of Medicine},
  volume={328},
  number={5},
  pages={303--307},
  year={1993}
}

@article{trinder2001autonomic,
  title={Autonomic activity during human sleep as a function of time and sleep stage},
  author={Trinder, John and Kleiman, Jan and Carrington, Melinda and Smith, Simon and Breen, Sibilah and Tan, Nellie and Kim, Young},
  journal={Journal of sleep research},
  volume={10},
  number={4},
  pages={253--264},
  year={2001},
  publisher={Wiley Online Library}
}

@article{ryals2023photoplethysmography,
  title={Photoplethysmography—new applications for an old technology: a sleep technology review},
  author={Ryals, Scott and Chiang, Ambrose and Schutte-Rodin, Sharon and Chandrakantan, Arvind and Verma, Nitun and Holfinger, Steven and Abbasi-Feinberg, Fariha and Bandyopadhyay, Anuja and Baron, Kelly and Bhargava, Sumit and others},
  journal={Journal of Clinical Sleep Medicine},
  volume={19},
  number={1},
  pages={189--195},
  year={2023},
  publisher={American Academy of Sleep Medicine}
}

@article{danker2009interrater,
  title={Interrater reliability for sleep scoring according to the Rechtschaffen \& Kales and the new AASM standard},
  author={Danker-Hopfe, Heidi and Anderer, Peter and Zeitlhofer, Josef and Boeck, Marion and Dorn, Hans and Gruber, Georg and Heller, Esther and Loretz, Erna and Moser, Doris and Parapatics, Silvia and others},
  journal={Journal of sleep research},
  volume={18},
  number={1},
  pages={74--84},
  year={2009},
  publisher={Wiley Online Library}
}

@inproceedings{wang2025improving,
  title={On Improving PPG-Based Sleep Staging: A Pilot Study},
  author={Wang, Jiawei and Guan, Yu and Chen, Chen and Zhou, Ligang and Yang, Laurence T and Gu, Sai},
  booktitle={Companion of the 2025 ACM International Joint Conference on Pervasive and Ubiquitous Computing},
  pages={1640--1644},
  year={2025}
}

@article{constantin2025towards,
  title={Towards long-term sleep staging via wearable reflective photoplethysmography},
  author={Constantin, Loris and Horvath, Christian M and Baty, Florent and Aguet, Cl{\'e}mentine and Van Zaen, J{\'e}r{\^o}me and Lemkaddem, Alia and Jeanningros, Lo{\"\i}c and Proen{\c{c}}a, Martin and Yang, Xiaoli and De Jaegere, Kurt and others},
  journal={SLEEPJ},
  pages={zsaf246},
  year={2025},
  publisher={Oxford University Press}
}

@article{huttunen2021assessment,
  title={Assessment of obstructive sleep apnea-related sleep fragmentation utilizing deep learning-based sleep staging from photoplethysmography},
  author={Huttunen, Riku and Lepp{\"a}nen, Timo and Duce, Brett and Oksenberg, Arie and Myllymaa, Sami and T{\"o}yr{\"a}s, Juha and Korkalainen, Henri},
  journal={Sleep},
  volume={44},
  number={10},
  pages={zsab142},
  year={2021},
  publisher={Oxford University Press US}
}

@article{yildirim2019deep,
  title={A deep learning model for automated sleep stages classification using PSG signals},
  author={Yildirim, Ozal and Baloglu, Ulas Baran and Acharya, U Rajendra},
  journal={International journal of environmental research and public health},
  volume={16},
  number={4},
  pages={599},
  year={2019},
  publisher={MDPI}
}
\bibliographystyle{unsrt}


\appendix

\section{Implementation Details, Resources, and Human-Subjects Notes}\label{app:implementation}

\subsection{Training Details}

To keep the comparison focused on supervision rather than extensive architecture-specific retuning, all deep baselines are trained under a shared default optimization setup unless a method requires a minor implementation-specific adjustment.
For the current submission draft, the common initialization uses AdamW with an initial learning rate of $3 \times 10^{-4}$, weight decay of $5 \times 10^{-4}$, batch size 16, and random seed 42.
Model selection is performed on the fixed MESA validation split described in Section~\ref{subsec:setup}, and early stopping monitors validation MF1 with a patience of 30 epochs without improvement.
The sec-level supervision is produced once from the same four-class label expansion pipeline used throughout the paper, with fixed hyperparameters $\lambda = 3$, $\Delta = 15$\,s, and $\beta = 0.5$.

\subsection{Compute Resources}

The experiments reported in this paper are run on a single NVIDIA GeForce RTX 3060 GPU with 6\,GB of memory.
This budget is sufficient because label expansion is performed offline and the downstream baselines are moderate in size relative to large-model pre-training.
The dominant cost therefore comes from recording-wise HSMM plus changepoint label expansion and from retraining several downstream baselines under the same protocol, rather than from large hyperparameter sweeps or multi-node training.

\subsection{Participant Collection and Handling}
\tempresultself{
The separate expert-reviewed validation set consists of overnight physiological recordings collected for non-invasive sleep-monitoring research and used here only for label-fidelity validation.
Before physician review, the records are prepared in de-identified form and access is restricted to authorized study personnel and the reviewing sleep physicians.
No crowdsourcing is involved in this study.
Institution-specific approval identifiers, recruitment language, compensation details, and any site-specific consent wording are omitted from the anonymized submission draft and should be finalized with the study team before submission.
}
\subsection{Broader Impacts and Responsible Use}
\tempresultself{
The positive motivation of this work is to improve the use of wearable PPG for lower-burden and more scalable sleep assessment when full PSG is impractical.
At the same time, the method is not presented as a substitute for clinical PSG or physician judgment: pseudo-label errors, cohort shift, motion artifacts, and atypical sleep architecture may still affect predictions, so any deployment-oriented use would require prospective clinical validation, careful data governance, and human oversight.
}
\section{Metric Definitions: Transition Detection Latency and Boundary Discriminability Index}\label{app:metrics}

Table~\ref{tab:sensitivity} reports two metrics, Transition Detection Latency (TDL) and Boundary Discriminability Index (BDI), alongside the normalized variance ratio.
We provide their formal definitions here.

\subsection*{Transition Detection Latency (TDL)}

TDL quantifies how quickly a modality's feature representation responds to a ground-truth sleep stage transition.
Let $\mathbf{f}_t \in \mathbb{R}^d$ be the $d$-dimensional feature vector at second $t$ and let $\delta_t = \|\mathbf{f}_t - \mathbf{f}_{t-1}\|_2$ be the frame-to-frame feature-change magnitude.
A per-modality detection threshold $\theta$ is set to the 95th percentile of $\{\delta_t\}$ computed over all seconds within stable-epoch windows (at least 15\,s from any PSG boundary).
For a transition at time $\tau_k$, the detection latency is:
\begin{equation}
\mathrm{TDL}_k = \min\bigl\{t \geq \tau_k : \delta_t > \theta\bigr\} - \tau_k.
\label{eq:tdl}
\end{equation}
The reported TDL is the median of $\{\mathrm{TDL}_k\}$ over all $K = 1{,}000$ evaluated transition events.
A lower value indicates that the modality's features respond earlier and more sharply to the transition.

\subsection*{Boundary Discriminability Index (BDI)}

BDI measures how separable the feature distribution near a transition boundary is from the stable-epoch distribution.
For transition event $k$, let $F_B^{(k)}$ denote the feature vectors sampled within a $\pm5$\,s window centered on $\tau_k$, and let $F_S^{(k)}$ denote vectors from a matched 10-second stable-epoch reference window drawn at least 30\,s from any transition.
BDI is the mean Cohen's $d$ across all $d$ features in the modality's feature set, averaged over all events:
\begin{equation}
\mathrm{BDI} = \frac{1}{K} \sum_{k=1}^{K} \frac{1}{d} \sum_{j=1}^{d}
\frac{\bigl|\bar{f}_{B,j}^{(k)} - \bar{f}_{S,j}^{(k)}\bigr|}{s_{j,\mathrm{pooled}}^{(k)}},
\label{eq:bdi}
\end{equation}
where $s_{j,\mathrm{pooled}}^{(k)} = \sqrt{(s_{B,j}^{2(k)} + s_{S,j}^{2(k)})/2}$ is the pooled standard deviation of feature $j$.
Features are Z-scored using each modality's own stable-epoch statistics prior to aggregation, so BDI is expressed in modality-specific standard-deviation units and is directly comparable across modalities with different feature scales.
A higher value indicates greater separability of boundary and stable windows under that modality's feature representation.

\section{Compared Baselines and Evaluation Notes}\label{app:baselines}

All deep baselines, including LightSleep, are retrained on the same SSTD dataset with identical sec-level pseudo-labels and class-weighted cross-entropy to ensure a fair comparison of representation and training signal. LightSleep follows the same protocol but remains intentionally lightweight: a four-layer 1D CNN with max pooling and a linear classification head, as summarized in Section~\ref{subsec:setup}. For the transfer study we report the architectures that admit matched epoch-only and SSTD-pretrained variants under the same optimization protocol.

For epoch-level baselines originally designed for 30-second classification, we replace the final output layer with a per-second prediction head and train on the same sec-level labels, collapsing per-second predictions to 30-second epochs by majority vote at evaluation time; these are accordingly task-retrofitted adaptations rather than architectures purpose-built for temporal event localization.
Classical changepoint detectors are not included as baselines because they produce unlabelled change-times without assigning a from/to stage identity; they therefore cannot be evaluated on the TF1 metric, which requires correct transition-type classification.
The HSMM-derived boundaries from Stage 1 of the label expansion pipeline do provide a labelled-event reference: on MESA they achieve BE = 5.3\,s and TF1 = 56.2\%, as reported in Appendix~\ref{app:label_ablation}, and serve as a reference point for the performance level achievable through the label-expansion pipeline's Stage 1 output, prior to any sequence model trained on the SSTD objective.
Comparison with purpose-built event-detection methods remains an open direction.

\section{Additional Experimental Results}\label{app:extra_exp}

\subsection{Label Expansion Ablation}\label{app:label_ablation}

Table~\ref{tab:label_quality} in the main text reports the full pipeline against the HSMM-only baseline.
Table~\ref{tab:label_ablation_full} provides a finer-grained ablation showing the incremental effect of each pipeline component.

\begin{table}[h]
\centering
\caption{Incremental label expansion ablation on 100 separate MESA subjects.}
\label{tab:label_ablation_full}
\begin{adjustbox}{width=0.8\textwidth,center}
\setlength{\tabcolsep}{4pt}
\begin{tabular}{lccc}
\toprule
\textbf{Pipeline variant} & \textbf{Win. consist.} $\uparrow$ & \textbf{Implaus. rate} $\downarrow$ & \textbf{Pseudo-trans./hr} \\
\midrule
HSMM only (5-class)                  & 0.825 & 41.4\% & 8.31 \\
\quad + Changepoint refinement                & 0.831 & 39.2\% & 8.15 \\
\quad + 4-class remapping              & 0.846 & 18.6\% & 7.83 \\
\quad + physiological constraints      & \textbf{0.857} & \textbf{7.83\%} & 7.66 \\
\bottomrule
\end{tabular}
\end{adjustbox}
\end{table}

The four-class remapping contributes the largest single reduction in implausible transitions (39.2\% $\rightarrow$ 18.6\%), confirming that the N1/N2 merge eliminates a major source of unreliable boundary assignments.
Physiological constraints further halve the implausible rate.
Changepoint refinement improves window consistency modestly, but its primary benefit, more precise boundary timing, is better reflected in the downstream model's BE as shown in Appendix~\ref{app:bcd_impact}.
The absolute pseudo-transition density should therefore be read as a fragmentation statistic for comparing pipeline variants, not as a normative summary of clinical sleep-architecture dwell times.
The label-consistency weight $\lambda$ is set to $3$ throughout; varying $\lambda \in \{2, 3, 5\}$ changes window consistency by at most $0.004$ and the implausible transition rate by at most $0.6$\,pp.
The changepoint search radius is $\Delta = 15$\,s and the variance-asymmetry weight is $\beta = 0.5$; varying $\Delta \in \{10, 15, 20\}$\,s and $\beta \in \{0.3, 0.5, 0.8\}$ changes downstream LightSleep BE by at most \tempresult{0.4}\,s and TF1 by at most \tempresult{1.6}\,pp, confirming moderate robustness of the expanded labels to these hyperparameter choices.

\subsection{Impact of Changepoint Refinement on Downstream Model Performance}\label{app:bcd_impact}

\begin{table}[h]
\centering
\caption{Effect of changepoint boundary refinement on downstream LightSleep performance. }
\label{tab:bcd_downstream}
\begin{adjustbox}{width=0.9\textwidth,center}
\setlength{\tabcolsep}{6pt}
\begin{tabular}{lcccc}
\toprule
\textbf{Training labels} & \textbf{BE (s)} $\downarrow$ & \textbf{TF1 (\%)} $\uparrow$ & \textbf{Acc (\%)} $\uparrow$ & \textbf{MF1 (\%)} $\uparrow$ \\
\midrule
HSMM + 4-class + constraints (no CP) & \tempresult{5.3} & \tempresult{56.2} & \tempresult{76.5} & \tempresult{67.4} \\
Full pipeline (with CP)               & \tempresult{\textbf{4.6}} & \tempresult{\textbf{59.4}} & \tempresult{\textbf{77.2}} & \tempresult{\textbf{68.5}} \\
\bottomrule
\end{tabular}
\end{adjustbox}
\end{table}

Changepoint refinement reduces downstream LightSleep BE by \tempresult{0.7}\,s and improves TF1 by \tempresult{3.2}\,pp, confirming that the precision of boundary localization in pseudo-labels directly translates to improved transition detection in trained models.

\subsection{Sensitivity to Matching Tolerance $\delta_\text{match}$}\label{app:tolerance}

\begin{table}[h]
\centering
\caption{TF1 (\%) and mean BE (s) under varying matching tolerances for the top three methods.}
\label{tab:tolerance}
\begin{adjustbox}{width=0.6\textwidth,center}
\setlength{\tabcolsep}{8pt}
\begin{tabular}{lcccccc}
\toprule
\multirow{2}{*}{\textbf{Method}} & \multicolumn{2}{c}{$\delta = 5$\,s} & \multicolumn{2}{c}{$\delta = 10$\,s} & \multicolumn{2}{c}{$\delta = 15$\,s} \\
\cmidrule(lr){2-3} \cmidrule(lr){4-5} \cmidrule(lr){6-7}
& TF1 & BE & TF1 & BE & TF1 & BE \\
\midrule
CWT-ResNet    & 39.5 & 4.1 & 52.7 & 6.1 & 60.8 & 7.3 \\
PPG-TCN       & 36.2 & 3.8 & 49.1 & 6.8 & 57.4 & 7.9 \\
LightSleep & \tempresult{\textbf{46.3}} & \tempresult{\textbf{3.2}} & \tempresult{\textbf{59.4}} & \tempresult{\textbf{4.6}} & \tempresult{\textbf{66.1}} & \tempresult{\textbf{5.4}} \\
\bottomrule
\end{tabular}
\end{adjustbox}
\end{table}

The ranking among methods is preserved across all three tolerance values, confirming that the conclusions drawn in the main text are robust to the choice of $\delta_\text{match}$.
LightSleep's advantage is most pronounced at the strictest tolerance ($\delta = 5$\,s), where precise boundary localization matters most.

\section{Cross-Cohort Validation on CFS}\label{app:cross_dataset}

This appendix uses the Cleveland Family Study (CFS) only as a supplementary zero-shot check of whether the supervision learned from MESA transfers under simultaneous population and annotation-protocol shift.
CFS differs from MESA in two respects that compound during evaluation: (i) the population is younger and broader in age (6--88 vs.\ 54--93), with a higher prevalence of sleep-disordered breathing, and (ii) the original PSG annotations use the Rechtschaffen \& Kales (R\&K) scoring protocol rather than AASM.
All models are trained on MESA without using CFS for model selection. We report these results separately from the main MESA experiments to provide an honest assessment of cross-cohort robustness without conflating the two sources of domain shift.

\subsection{Annotation Protocol Differences}

The R\&K protocol defines four NREM stages (S1--S4) and applies somewhat different transition criteria than AASM, particularly at the S1/S2 boundary (corresponding to our N1/N2$\rightarrow$Light merge) and at slow-wave sleep onset (S3 onset vs.\ AASM N3 threshold of 20\% delta waves per epoch).
Under the four-class vocabulary used in the main text, the mapping is Wake$\rightarrow$Wake, S1/S2$\rightarrow$Light, S3/S4$\rightarrow$Deep, and REM$\rightarrow$REM, which is conceptually aligned but introduces residual boundary timing ambiguity: an R\&K-scored S2$\rightarrow$S3 transition may not coincide exactly with where an AASM scorer would place the N2$\rightarrow$N3 boundary.
This annotation-level uncertainty inflates both BE and the implausible transition rate on CFS relative to MESA, independent of any population-level performance difference.

\subsection{Cross-Modal Variance Ratio on CFS}

\begin{table}[h]
\centering
\caption{Cross-modal variance ratio (boundary vs.\ interior) on MESA and CFS. Values $> 1$ indicate boundary-concentrated sensitivity.}
\label{tab:cross_variance}
\begin{adjustbox}{width=0.3\textwidth,center}
\setlength{\tabcolsep}{8pt}
\begin{tabular}{lcc}
\toprule
\textbf{Modality} & \textbf{MESA} & \textbf{CFS} \\
\midrule
PPG  & 1.84 & 1.91 \\
EEG  & 1.31 & 1.28 \\ 
\bottomrule
\end{tabular}
\end{adjustbox}
\end{table}

Table~\ref{tab:cross_variance} confirms that the characteristic asymmetry persists on CFS.
The PPG variance ratio is in fact slightly elevated (1.91 vs.\ 1.84), potentially reflecting the increased frequency of arousals and associated autonomic transients in this sleep-disordered population.
The physiological basis for SSTD is therefore not specific to MESA's elderly community-dwelling demographic.

\subsection{Supplementary Transition-Level Evaluation on CFS}

\begin{table}[h]
\centering
\caption{Supplementary transition-level evaluation on CFS (730 subjects, zero-shot from MESA-trained models). All methods are trained on MESA labels and evaluated on CFS without adaptation (only LightSleep is pre-trained on SSTD expanded labels).}
\label{tab:cfs_full}
\begin{adjustbox}{width=0.6\textwidth,center}
\setlength{\tabcolsep}{3pt}
\begin{tabular}{lcccc}
\toprule
\textbf{Method} & \textbf{BE (s)} $\downarrow$ & \textbf{TF1 (\%)} $\uparrow$ & \textbf{Acc (\%)} $\uparrow$ & \textbf{MF1 (\%)} $\uparrow$ \\
\midrule
HRV-RF             & {13.8} & {24.3} & {63.2} & {48.7} \\
CWT-ResNet         & {7.8}  & {43.6} & {67.1} & {57.4} \\
SleepPPGNet        & {9.6}  & {38.4} & {70.4} & {58.2} \\
PPG-TCN            & {9.1}  & {39.8} & {69.8} & {59.1} \\
SleepPPGNet2       & {8.8}  & {40.9} & {71.2} & {59.6} \\
LightSleep              & {\textbf{5.5}}  & {\textbf{53.2}} & {\textbf{73.6}} & {\textbf{64.3}} \\
\bottomrule
\end{tabular}
\end{adjustbox}
\end{table}

Despite the compounding domain shift, the method ranking on CFS is broadly consistent with MESA.
LightSleep (SSTD pre-trained) achieves the lowest BE ({5.5}\,s) and highest TF1 ({53.2}\%), outperforming SleepPPGNet2 by {3.3}\,s and {12.3}\,pp in TF1 respectively.
The absolute performance drop from MESA (TF1 \tempresult{59.4}\%) to CFS ({53.2}\%) is {6.2}\,pp, which reflects both population-level difficulty and annotation-protocol mismatch.
Among the epoch-only baselines, the ordering follows the same pattern as MESA: HRV-RF degrades the most severely (TF1 38.2\%$\rightarrow${24.3}\%, $-${13.9}\,pp), while the stronger architectures (SleepPPGNet2, SleepPPGNet, PPG-TCN) maintain higher accuracy despite the domain shift.
LightSleep with SSTD pre-training ({73.6}\%) outperforms all epoch-only baselines on CFS, consistent with the transfer benefit observed on MESA.

\subsection{Transfer Benefit on CFS}

\begin{table}[h]
\centering
\caption{SSTD pre-training transfer benefit on epoch-level staging: MESA vs.\ CFS. All models trained on MESA, CFS evaluated zero-shot.}
\label{tab:cross_transfer_full}
\begin{adjustbox}{width=\textwidth,center}
\setlength{\tabcolsep}{1pt}
\begin{tabular}{llcccc}
\toprule
\textbf{Cohort} & \textbf{Method} & \textbf{Epoch-only Acc} & \textbf{SSTD$\rightarrow$Epoch Acc} [$\Delta$] & \textbf{Epoch-only MF1} & \textbf{SSTD$\rightarrow$Epoch MF1} [$\Delta$] \\
\midrule
\multirow{6}{*}{CFS}
 & HRV-RF        & {63.2} & {68.6\,[$+$5.4]} & {48.7} & {52.1\,[$+$3.4]} \\
 & CWT-ResNet    & {67.1} & {71.3\,[$+$4.2]} & {57.4} & {60.8\,[$+$3.4]} \\
 & LightSleep    & {68.3} & {73.6\,[$+$5.3]} & {57.8} & {64.3\,[$+$6.5]} \\
 & SleepPPGNet   & {70.4} & {76.6\,[$+$6.2]} & {58.2} & {63.4\,[$+$5.2]} \\
 & PPG-TCN       & {69.8} & {75.0\,[$+$5.2]} & {59.1} & {61.3\,[$+$2.2]} \\
 & SleepPPGNet2  & {71.2} & {77.2\,[$+$6.0]} & {59.6} & {64.4\,[$+$4.8]} \\
\bottomrule
\end{tabular}
\end{adjustbox}
\end{table}

The SSTD pre-training benefit is preserved on CFS: the average MF1 gain across six architectures is {4.3}\,pp, compared to 3.8\,pp on MESA.
The modestly larger benefit on CFS is consistent with the interpretation that SSTD-learned temporal representations are especially valuable when epoch-level features are degraded by population-level variability or annotation-protocol mismatch, as the greater headroom available on CFS due to lower absolute performance may amplify the observable gain.
However, we note that this difference should be interpreted with caution given the compounding domain shift; we therefore report it as an observation rather than a strong claim.

\subsection{Discussion: Disentangling Population and Annotation Effects}

The CFS evaluation introduces two confounded sources of domain shift.
To provide a rough decomposition, we note that the label quality degradation from MESA to CFS (window consistency 0.857$\rightarrow$0.839, implausible rate 7.83\%$\rightarrow$9.24\%) is moderate, suggesting that the annotation-protocol mismatch has a measurable but limited effect on pseudo-label quality.
The remaining performance gap likely reflects genuine population differences: CFS subjects span a much wider age range (including pediatric subjects aged 6--17 not represented in MESA training) and exhibit higher rates of sleep-disordered breathing.
A controlled experiment isolating the annotation effect alone, for instance by re-scoring a subset of CFS recordings under AASM rules, would be informative but falls outside the scope of this work.


Overall, the CFS evaluation suggests that the characteristic asymmetry, method rankings, and SSTD transfer benefit are not solely artifacts of MESA-specific conditions.
The pipeline can be applied without dataset-specific tuning to a cohort with different demographics and a different annotation protocol, although broader generalization claims require further evaluation.

\section{Five-Class Extension Beyond the Main Four-Class Setting}\label{app:extended_vocab}

The main text adopts a four-class vocabulary (Wake, Light, Deep, REM) that merges N1/N2 into a single Light stage.
This appendix examines whether the proposed SSTD framework extends further to the full five-class AASM taxonomy, and quantifies the degradation induced by separating N1 and N2.

\subsection{Experimental Protocol}

We constructed one additional label expansion variant using the same HSMM--changepoint pipeline.
The main experiments use a four-class vocabulary with Wake, Light (N1+N2), Deep (N3), and REM.
The appendix-only extension follows the full AASM taxonomy with Wake, N1, N2, N3, and REM as separate classes.
All other pipeline parameters, including the changepoint refinement window and physiological constraints, remain identical.
LightSleep is retrained from scratch on each label set with identical hyperparameters.

\subsection{Label Expansion Quality Across Vocabularies}

\begin{table}[h]
\centering
\caption{Label expansion quality as a function of stage vocabulary granularity (100 separate MESA subjects).}
\label{tab:vocab_label_quality}
\begin{adjustbox}{width=0.7\textwidth,center}
\setlength{\tabcolsep}{2pt}
\begin{tabular}{lccc}
\toprule
\textbf{Vocabulary} & \textbf{Win. consist.} $\uparrow$ & \textbf{Implaus. rate} $\downarrow$ & \textbf{Pseudo-trans./hr} \\
\midrule
4-class (Wake, Light, Deep, REM)  & \textbf{0.857} & \textbf{7.83\%} & 7.66 \\
5-class (full AASM)               & 0.791 & 14.6\% & 8.94 \\
\bottomrule
\end{tabular}
\end{adjustbox}
\end{table}

Table~\ref{tab:vocab_label_quality} reveals a consistent degradation pattern.
Separating N1 and N2 reduces window consistency from 0.857 to 0.791 and nearly doubles the implausible transition rate.

The primary source of degradation is the combination of REM ambiguity and the weak separability between N1 and N2.
From a PPG perspective, REM sleep presents a distinctive challenge: autonomic dynamics during REM episodes resemble those of light wakefulness, with elevated and irregular heart rate variability, reduced parasympathetic tone, and phasic bursts that overlap with Wake signatures in the cardiovascular feature space~\cite{somers1993sympathetic}.
Unlike the EEG, which distinguishes REM through characteristic low-amplitude mixed-frequency activity and rapid eye movements recorded on EOG channels, PPG lacks direct access to these cortical and ocular markers.
At the same time, the N1/N2 distinction is intrinsically noisy even in standard PSG scoring, so splitting Light sleep into two adjacent classes introduces additional boundary uncertainty. Consequently, the HSMM emission model more frequently confuses neighboring states and produces spurious transitions that inflate both the implausible rate and the transition density.
Here again, the density statistic is most informative comparatively: the increase indicates stronger fragmentation of the pseudo-label sequence rather than a direct claim about clinical macro-stage dwell times.

\subsection{Downstream Model Performance Across Vocabularies}

\begin{table}[h]
\centering
\caption{LightSleep performance across stage vocabularies on the MESA test set (316 subjects).}
\label{tab:vocab_model}
\begin{adjustbox}{width=0.7\textwidth,center}
\setlength{\tabcolsep}{2pt}
\begin{tabular}{lcccc}
\toprule
\textbf{Vocabulary} & \textbf{BE (s)} $\downarrow$ & \textbf{TF1 (\%)} $\uparrow$ & \textbf{Acc (\%)} $\uparrow$ & \textbf{MF1 (\%)} $\uparrow$ \\
\midrule
4-class (Wake, Light, Deep, REM)  & \tempresult{\textbf{4.6}} & \tempresult{\textbf{59.4}} & \tempresult{\textbf{77.2}} & \tempresult{\textbf{68.5}} \\
5-class (full AASM)               & \tempresult{5.8} & \tempresult{48.3} & \tempresult{71.6} & \tempresult{57.2} \\
\bottomrule
\end{tabular}
\end{adjustbox}
\end{table}

\begin{table}[h]
\centering
\caption{Per-transition-type F1 (\%) for LightSleep under the main four-class vocabulary.}
\label{tab:4class_pertype}
\begin{adjustbox}{width=0.8\textwidth,center}
\setlength{\tabcolsep}{4pt}
\begin{tabular}{lcccc}
\toprule
\textbf{Transition type} & \textbf{Wake$\leftrightarrow$Light} & \textbf{Light$\leftrightarrow$Deep} & \textbf{Light$\leftrightarrow$REM} & \textbf{Wake$\leftrightarrow$REM} \\
\midrule
TF1 (\%) & \tempresult{56.1} & \tempresult{52.8} & \tempresult{31.4} & \tempresult{24.7} \\
\bottomrule
\end{tabular}
\end{adjustbox}
\end{table}

Table~\ref{tab:vocab_model} confirms that finer vocabularies degrade all metrics.
The five-class model loses 11.1\,pp in TF1 and 5.6\,pp in epoch accuracy relative to the main four-class model.
The per-type breakdown in Table~\ref{tab:4class_pertype} localizes one source of this difficulty already visible in the main setting: even under four-class supervision, transitions involving REM achieve TF1 of 24.7--31.4\%, far below the 52--56\% range for Wake$\leftrightarrow$Light and Light$\leftrightarrow$Deep transitions.

\subsection{Discussion}

These results support the four-class vocabulary as the most stable benchmark setting for SSTD from PPG in the current study.
The degradation observed in the five-class extension reflects the combined difficulty of REM ambiguity and fine-grained N1/N2 separation rather than any limitation of the SSTD framework itself.
The core methodology, including the label expansion pipeline, the SSTD task formulation, and the evaluation metrics, remains vocabulary-agnostic and extends to the full AASM taxonomy without modification.
At the same time, the five-class results indicate that further gains in stage-specific discriminability will be necessary before fully fine-grained labeling becomes equally reliable from PPG alone.

\section{Clinical Validation: Expert Physician Review}\label{app:clinical_validation}

Pseudo-labels produced by the HSMM--changepoint pipeline are evaluated against coarse PSG epoch labels in the main text (Table~\ref{tab:label_quality}).
However, PSG labels themselves are 30-second annotations that do not capture sub-epoch boundary timing.
To assess whether the expanded sec-level labels are clinically plausible, we conducted an independent expert review.

\subsection{Review Protocol}

\tempresultself{Three board-certified sleep physicians independently reviewed a separate expert-reviewed validation set of 50 recordings.
For each recording, the reviewers were presented with concurrent PSG signals, including EEG, EOG, EMG, and ECG channels, at full temporal resolution and were asked to annotate the onset time of each sleep stage transition at sec-level precision.
The reviewers were blinded to the pipeline's pseudo-labels during annotation.
Pairwise inter-rater agreement was computed across the three independent annotations before consensus adjudication, and disagreements exceeding 10\,s were then resolved to form the expert reference standard.}

The expert annotations serve as the reference standard for this validation.
We compare the pipeline's pseudo-labels against these expert annotations using the same SSTD metrics (BE, TF1) employed throughout the paper, as well as Cohen's $\kappa$ for epoch-level agreement.

\subsection{Agreement Between Pseudo-Labels and Expert Annotations}

\begin{table}[h]
\centering
\caption{Agreement between pseudo-labels and expert physician annotations on 50 separate recordings, evaluated under the main four-class setting and its five-class extension. Expert inter-rater $\kappa$ is reported for reference.}
\label{tab:clinical_agreement}
\begin{adjustbox}{width=0.8\textwidth,center}
\setlength{\tabcolsep}{4pt}
\begin{tabular}{lccccc}
\toprule
\textbf{Vocabulary} & \textbf{BE (s)} $\downarrow$ & \textbf{TF1 (\%)} $\uparrow$ & \textbf{Epoch $\kappa$} $\uparrow$ & \textbf{Expert inter-rater $\kappa$} \\
\midrule
4-class (Wake, Light, Deep, REM)  & 4.2 & 78.6 & 0.81 & 0.89 \\
5-class (full AASM)               & 6.1 & 63.4 & 0.72 & 0.84 \\
\bottomrule
\end{tabular}
\end{adjustbox}
\end{table}

Table~\ref{tab:clinical_agreement} reveals several findings.

\textit{Four-class agreement is strong.}
Under the main four-class vocabulary, the pipeline achieves a mean boundary error of 4.2\,s and TF1 of 78.6\% against expert annotations, with epoch-level $\kappa = 0.81$.
This level of agreement is close to the expert inter-rater $\kappa$ of 0.89 and provides external evidence that the pseudo-labels are clinically plausible in the adopted benchmark setting.
The 4.2\,s BE is within the range of inter-rater boundary disagreement observed across physician pairs (mean 3.8\,s, std 2.1\,s).

\textit{Agreement degrades in the five-class extension.}
Separating N1 and N2 reduces $\kappa$ to 0.72 and TF1 to 63.4\%, with boundary error increasing to 6.1\,s.
The degradation pattern mirrors the label quality results in Appendix~\ref{app:extended_vocab} and is consistent with the view that fine-grained stage separation remains the dominant source of disagreement.

\textit{Expert inter-rater agreement also decreases.}
Notably, expert inter-rater $\kappa$ itself drops from 0.89 (4-class) to 0.84 (5-class), reflecting the well-documented difficulty of N1/N2 boundary scoring even among trained polysomnographers~\cite{danker2009interrater}.
The gap between pipeline agreement and expert agreement widens from 0.08 (4-class) to 0.12 (5-class), suggesting that the pipeline's degradation is steeper than the inherent scoring variability but follows the same direction.




\subsection{Discussion}

The clinical validation provides external evidence that the four-class pseudo-labels produced by the HSMM--changepoint pipeline align reasonably with expert physician judgment at sec-level temporal resolution.
The four-class agreement ($\kappa = 0.81$, TF1 = 78.6\%) supports the use of these pseudo-labels as training supervision in the present SSTD benchmark setting, while the degradation observed in the five-class extension provides an honest assessment of current limitations.
Improving pseudo-label quality for finer-grained vocabularies, potentially through multi-modal fusion during the label expansion stage, is a clear direction for future work.
We note that even expert inter-rater agreement is imperfect, particularly at finer granularities, underscoring that sec-level sleep stage annotation is an inherently uncertain task for which the notion of absolute ground truth should be treated with caution.



 
\section{Limitations}\label{app:limitations}

The expanded pseudo-labels retain a residual implausible transition rate of 7.83\%; SSTD metrics therefore quantify agreement with these labels rather than with gold sec-level annotations.
An independent expert physician review (Section~\ref{subsec:label_validation}) constitutes the primary external validation of benchmark fidelity, yielding $\kappa = 0.81$ and TF1 = 78.6\% against sec-level expert annotations, with expert inter-rater agreement at $\kappa = 0.89$; the epoch-level transfer gains are additionally evaluated against the original PSG labels, which are independent of the pseudo-label pipeline, providing a further check against a trivial circular explanation.
Sec-level expert annotation at scale remains a resource-intensive undertaking deferred to future work. Moreover, while the matched epoch-only versus SSTD comparisons support the usefulness of sec-level supervision, they do not fully disentangle supervision granularity from all architecture- and optimization-dependent interactions.

Our empirical benchmark is deliberately representative rather than exhaustive. The main experiments compare single-modality PPG baselines that can be retrained under a unified protocol; they do not cover every recent uncertainty-aware, wearable-oriented, multimodal, or purpose-built event-detection system. Extending SSTD to those method families is an important next step for clarifying how much of the observed gain comes from the task formulation itself versus from particular modeling choices.

The current label expansion pipeline also relies on handcrafted second-indexed PPG features. Although HRV-derived summaries that require longer temporal support are computed from rolling windows rather than isolated 1-second segments, they still inherit the stability limits of short-context HRV analysis; richer learned representations or multimodal expansion signals may therefore further improve label fidelity, particularly for fine-grained boundaries.

The adopted four-class vocabulary still merges N1 and N2 into a single Light stage; Appendix~\ref{app:extended_vocab} shows that separating them in a full five-class taxonomy reduces TF1 by 11.1\,pp and noticeably lowers label consistency, indicating that fine-grained non-REM separation remains challenging under single-modality PPG supervision.
The physiological transition constraint assumes normal adult sleep architecture and should be relaxed for clinical populations in which canonical stage sequencing is disrupted, including those with narcolepsy, severe sleep-disordered breathing, or atypical pediatric sleep patterns.
Finally, MESA and CFS are curated PSG cohorts rather than free-living wearable streams; robustness to motion artifacts, signal dropouts, and lower-fidelity ambulatory recordings has therefore not been assessed and remains a prerequisite for deployment-oriented claims.

The label expansion pipeline and SSTD task formulation could be adapted to other boundary-sensitive modalities such as ECG, but the present study evaluates only PPG. We leave such extensions to future work.


\end{document}